\documentclass[12pt]{article} 
\usepackage{bbold}
\usepackage{cite}
\usepackage{graphicx}
\def \bea{\begin{eqnarray}} 
\def \beq{\begin{equation}}

\def \eea{\end{eqnarray}} 
\def \eeq{\end{equation}}

\def \half{\frac{1}{2}}

\def\lsim{\mathrel{\rlap{\lower3pt\hbox{$\sim$}}\raise2pt\hbox{$<$}}}
\def\gsim{\mathrel{\rlap{\lower3pt\hbox{$\sim$}}\raise2pt\hbox{$>$}}}

\begin{document} 
\begin{flushright}
TECHNION-PH-2016-17 \\
EFI 16-27 \\
December 2016 \\
\end{flushright} 
\centerline{\bf Controlling $\rho$ width effects for a precise value of
$\alpha$ in $B \to \rho \rho$}
\medskip
\centerline{Michael Gronau}
\centerline{\it Physics Department, Technion, Haifa 32000, Israel}
\medskip 
\centerline{Jonathan L. Rosner} 
\centerline{\it Enrico Fermi Institute and Department of Physics,
  University of Chicago} 
\centerline{\it Chicago, IL 60637, U.S.A.} 
\bigskip

\begin{quote}
It has been pointed out that the currently most precise determination of the
weak phase $\phi_2 = \alpha$ of the Cabibbo-Kobayashi-Maskawa (CKM) matrix
achieved in $B \to \rho\rho$ decays is susceptible to a small correction at a
level of $(\Gamma_\rho/m_\rho)^2$ due to an $I=1$ amplitude caused by the
$\rho$ width.  Using Breit-Wigner distributions for the two pairs of pions
forming $\rho$ mesons, we study the $I=1$ contribution to $B\to \rho\rho$ decay
rates as function of the width and location of the $\rho$ band. We find that in
the absence of a particular enhancement of the $I=1$ amplitude reducing a
single band to a width $\Gamma_\rho$ at SuperKEKB leads to results which are
completely insensitive to the $\rho$ width.  If the $I=1$ amplitude is
dynamically enhanced relative to the $I=0,2$ amplitude one could subject its
contribution to a ``magnifying glass'' measurement using two
separated $\rho$ bands of width $\Gamma_\rho$.  Subtraction of the $I=1$
contribution from the measured decay rate would lead to a very precise
determination of the $I=0,2$ amplitude needed for performing the isospin
analysis.
\end{quote}

\leftline{\qquad PACS codes: 12.15.Hh, 13.25.Hw, 14.40.Nd}
\bigskip

\section{Introduction}

Precision measurements of phases in the Cabibbo-Kobayashi-Maskawa (CKM) matrix
responsible for CP violation are one way for detecting new physics. The most
accurate method for determining $\phi_2 = \alpha \equiv {\rm Arg}(-V^*_{tb}
V_{td}/V^*_{ub}V_{ud})$ is based on $B \to \rho\rho$ decays. Isospin symmetry
implies amplitude triangle relations for longitudinally polarized $\rho$ mesons
in $B^0 \to \rho^+\rho^-, \rho^0\rho^0$ and $B ^+ \to \rho^+\rho^0$ and their
charge-conjugates, which specify a value for $\alpha$ when being augmented by
time-dependent CP asymmetries in the first two processes~\cite{Gronau:1990ka}.
The current status of applying this method has been described recently in Ref.\,
\cite{Gronau:2016idx}, leading to a present error of $5^\circ$ in $\alpha$,
and projecting an error less than one degree for future experiments to be performed 
by the Belle II Collaboration at SuperKEKB~\cite{Abe:2010gxa,Aushev:2010bq}. 

In the above method one neglects the $\rho$ width assuming equal masses for the
two final $\rho$ mesons, which by Bose symmetry must be in $I=0$ and $I=2$
states. The authors of Ref.\,\cite{Falk:2003uq} pointed out that the $\rho$
width introduces a new isospin amplitude, because two $\rho$ mesons observed
with different invariant masses may have a total $I=1$.  The $I=1$ contribution
to $B\to \rho\rho$ decay rates was suggested 
on dimensional grounds to decrease with the width $\Delta$ of the $\rho$ band
at least as $(\Delta/m_\rho)^2$ for $\Delta < \Gamma_\rho$. In order to eliminate 
corrections in $\alpha$ due to the $\rho$ width Ref.\,\cite{Falk:2003uq} 
proposed to measure decay rates for decreasing values of 
$\Delta$, reaching a point where these measurements become stable under variation 
of small values of $\Delta$ while paying a price in statistics.

The purpose of this Letter is to 
calculate explicitly the $I=1$ relative contributions in $B\to \rho\rho$ decay rates, 
studying in detail their dependence on the widths {\em and location} of the two $\rho$ 
meson bands. It will be shown that these contributions for a common band behave like 
$(\Delta/m_\rho)^2$ also for $\Delta > \Gamma_\rho$, with a calculable coefficient 
$\sim 0.1$ increasing moderately with decreasing $\Delta$, reaching an asymptotic 
behavior $\frac{1}{6}(\Delta/m_\rho)^2$ for $\Delta \ll 2\Gamma_\rho$.
As their absolute normalization is a priori unknown one must 
also consider the possibility that they are dynamically enhanced relative to the $I = 0, 2$ 
terms, in which case they would remain significant also for $\Delta = \Gamma_\rho$.  
A judicious choice of two different locations of the two $\rho$ meson bands will be
used to artificially enhance this $I=1$ contribution by a large factor relative to 
$(\Gamma_\rho/m_\rho)^2$. This enables resolving completely the uncertainty 
in $\alpha$ also when the $I=1$ amplitude is dynamically enhanced.

In Section 2 we study the contribution of an $I=1$ amplitude to $B \to
\rho\rho$ decay rates for two $\rho$ mesons in a common mass band centered at
$m_\rho$, assuming for each $\rho$ meson decay into two pions a Breit-Wigner
distribution. Similar calculations are performed in Sections 3 for two $\rho$
meson bands adjacent to each other and in Section 4 for two bands separated
from each other, lying in both cases above and below $m_\rho$. 
We summarize our results numerically and conclude in Section 5.

\section{Two $\rho$ mesons in a common band centered at $m_\rho$}

The following discussion applies separately to the decays $B^0 \to
\rho^+\rho^-$ and $B^+\to \rho^+\rho^0$, in which the final $\rho$ mesons have
been measured to be almost $100\%$ longitudinally polarized
\cite{Gronau:2016idx}. It does not apply to $B^0\to \rho^0\rho^0$ in which the
final state cannot be in $I=1$. We will consider decay amplitudes $A_L(B \to
\rho\rho)$ for longitudinally polarized $\rho$ mesons in the first two 
processes. These amplitudes depend on two variables, the two dipion invariant
masses, $m^2_{12} \equiv (p_1 + p_2)^2$ and $m^2_{34} \equiv (p_3 + p_4)^2$,
through Breit-Wigner distributions: 
\beq
A_L(B \to \rho\rho) = f(m_{12}, m_{34})
\left(\frac{m_\rho\Gamma_\rho}{m^2_{12} - m^2_\rho + im_\rho\Gamma_\rho}\right)
\left(\frac{m_\rho\Gamma_\rho}{m^2_{34} - m^2_\rho + im_\rho\Gamma_\rho}\right)~.
\eeq
The decay amplitude involves two parts corresponding to final states involving
isospin zero or two (two alone for $B^+ \to \rho^+\rho^0$) and isospin one,
respectively:
\beq
f(m_{12}, m_{34})  =  f_{I=0,2}(m_{12}, m_{34}) + f_{I=1}(m_{12}, m_{34})~.
\eeq
These two parts are symmetric and antisymmetric, respectively, under 
interchanging $m_{12}$ and $m_{34}$, 
\beq
f_{I=0,2}(m_{12}, m_{34})  =  f_{I=0,2}(m_{34}, m_{12})~,~~~~
f_{I=1}(m_{12}, m_{34}) = -f_{I=1}(m_{34}, m_{12})~,
\eeq
implying
\beq
f(m_{12},m_{34})  =  f_{I=0,2}(m_{12},m_{34}) - f_{I=1}(m_{34},m_{12})~.
\eeq

We will consider the longitudinal decay rate $\Gamma_L(B \to \rho\rho)$ for 
invariant masses $m_{12}, m_{34}$ lying in a common range $M$ symmetric 
about $m_\rho$:
\bea
\Gamma_L(B^0 \to \rho^+\rho^-)_M & = & (m_\rho\Gamma_\rho)^4
\int_M\int_M\frac{|f(m_{12}, m_{34})|^2
 dm_{12}dm_{34}}{|m^2_{12} - m^2_\rho + im_\rho\Gamma_\rho|^2
|m^2_{34} - m^2_\rho + im_\rho\Gamma_\rho|^2}
\\
& = &  (m_\rho\Gamma_\rho)^4
\int_M\int_M\frac{(|f_{I=0,2}(m_{12}, m_{34})|^2 + |f_{I=1}(m_{12}, m_{34})|^2)
 dm_{12} dm_{34}}{|m^2_{12} - m^2_\rho + im_\rho\Gamma_\rho|^2
|m^2_{34} - m^2_\rho + im_\rho\Gamma_\rho|^2}~.
\nonumber
\eea
The interference term is antisymmetric in $m_{12} \leftrightarrow m_{34}$ and vanishes when 
these two variables are integrated over a common range\,\cite{Falk:2003uq},
\beq
 \int_M\int_M\frac{{\rm Re}[f_{I=0,2}(m_{12}, m_{34})f_{I=1}^*(m_{12}, m_{34})]dm_{12} dm_{34}}
{ |m^2_{12} - m^2_\rho + im_\rho\Gamma_\rho|^2
|m^2_{34} - m^2_\rho + im_\rho\Gamma_\rho|^2} = 0~.
\eeq

The leading term in the $I=0,2$ amplitude behaves like a constant, $f_{I=0,2} = a$, 
while that of the $I=1$ amplitude behaves like $f_{I=1} = c(m_{12} - m_{34})/m_\rho$. 
The authors of Ref.\,\cite{Falk:2003uq} assume that the two constants $a$ and $c$ 
are of the same order. 
This seems like a reasonable assumption which needs to be 
tested experimentally. Thus the ratio $R_L$ of the contributions of these amplitudes to 
the decay rate is:
\bea\label{RL}
R_L & = &\frac{c^2}{a^2}\int_M\int_M \frac{[(m_{12} - m_{34})/m_\rho]^2
 dm_{12} dm_{34}}{[(m^2_{12} - m^2_\rho)^2 + (m_\rho\Gamma_\rho)^2] [m_{12} \to m_{34}]}
 \large/\left(\int_M\frac{dm}{(m^2 - m^2_\rho)^2 + (m_\rho\Gamma_\rho)^2}\right)^2
 \nonumber\\
 & = & \frac{c^2}{a^2}\frac{I_1}{(I_0)^2}~.
\eea
$I_1$ and $(I_0)^2$ denote the integrals in the numerator and denominator, characterizing 
(up to a ratio $c^2/a^2$) contributions of $I=1$ and $I=0,2$ amplitudes to $B\to \rho\rho$ 
decay rates. In this section and in the next two sections we will study their ratio for varying 
ranges of the two $\rho$ meson bands.

Defining $\gamma \equiv \Gamma_\rho/m_\rho,~x \equiv m/m_\rho,~x_1 
\equiv m_{12}/m_\rho,~x_2 \equiv m_{34}/m_\rho$,
$\delta \equiv \Delta/2m_\rho$, and
considering a range $1 - \delta \le x, x_1, x_2 \le 1 + \delta$ for these three variables
corresponding to a common mass band of width $\Delta$ for $m_{12}$ and $m_{34}$, 
$m_\rho(1 - \delta) \le m_{12}, m_{34} \le m_\rho(1 + \delta)$, 
we have
 \beq\label{I0}
I_0 \equiv \int_{1- \delta}^{1 + \delta}\frac{dx}{(x^2 - 1)^2 + \gamma^2}~,
\eeq
\bea\label{I1f}
I_1 & \equiv & \int_{1-\delta}^{1+\delta}\int_{1-\delta}^{1+\delta}\frac{(x_1-x_2)^2dx_1dx_2}
{[(x_1^2 - 1)^2 + \gamma^2][(x_2^2 - 1)^2 + \gamma^2]}
\nonumber\\
& = & 2\int_{1- \delta}^{1 + \delta}\frac{x^2dx}{(x^2 - 1)^2 + \gamma^2}
\int_{1- \delta}^{1 + \delta}\frac{dx}{(x^2 - 1)^2 + \gamma^2}
- 2\left(\int_{1- \delta}^{1 + \delta}\frac{xdx}{(x^2 - 1)^2 + \gamma^2}\right)^2~.
\eea

Thus we are interested in the following integrals,
\beq
J_0 \equiv \int_{1-\delta}^{1+\delta}\frac{dx}{(x^2 - 1)^2+\gamma^2}~,~~~
J_1 \equiv \int_{1-\delta}^{1+\delta}\frac{xdx}{(x^2 - 1)^2+\gamma^2}~,~~~
J_2 \equiv \int_{1-\delta}^{1+\delta}\frac{x^2dx}{(x^2 - 1)^2+\gamma^2}~,
\eeq
where we want to calculate $I_0 = J_0$, $I_1 = 2[J_2 J_0 - (J_1)^2]$, and
$I_1/(I_0)^2$.  We will study cases where $\gamma$ and
$\delta$ (of order $\gamma$ or smaller) are small in comparison with 1, so the main 
contributions to the integrals come from values of $x$ close to 1.

We substitute $u = x^2 -1$, so $du = 2xdx = 2\sqrt{1+u}~dx$.  For small $u$
the limits of integration $x = 1 \pm \delta$ translate to $u = \pm 2 \delta$,
and the integrals can be written
\beq
I_0 = \frac{1}{2}\int_{-2\delta}^{2\delta}\frac{(1 + u)^{-1/2}du}{u^2 + \gamma^2}~, 
\eeq
\beq
I_1 = \frac{1}{2}\int_{-2\delta}^{2\delta}\frac{(1+u)^{1/2}du}{u^2+\gamma^2}
\int_{-2\delta}^{2\delta}\frac{(1+u)^{-1/2}du}{u^2+\gamma^2} -\frac{1}{2}\left(
\int_{-2\delta}^{2\delta}\frac{du}{u^2+\gamma^2} \right)^2~.
\eeq
Applying two simple integral functions,
\beq\label{integrals}
\int^u\frac{du}{u^2 + \gamma^2} = \frac{1}{\gamma}{\rm arctan}\left(\frac{u}{\gamma}\right)~,~~~~
\int^u\frac{u^2du}{u^2 + \gamma^2} = \int^u\frac{(u^2 + \gamma^2 - \gamma^2)du}{u^2 + \gamma^2}
= u - \gamma\,{\rm arctan}\left(\frac{u}{\gamma}\right)~,
\eeq
implies
\beq
\int^{2\delta}_{-2\delta}\frac{du}{u^2 + \gamma^2} = \frac{2}{\gamma}
{\rm arctan}\left(\frac{2\delta}{\gamma}\right)~,~~~~
\int^{2\delta}_{-2\delta}\frac{u^2du}{u^2 + \gamma^2} = 
4\delta - 2\gamma\,{\rm arctan}\left(\frac{2\delta}{\gamma}\right)~.
\eeq

Using the binomial expansions $(1+u)^{1/2} = 1 + (u/2) - (u^2/8) + \ldots$ and
$(1+u)^{-1/2} = 1 - (u/2) + 3(u^2/8) - \ldots$, collecting terms, omitting
integrals whose integrands are odd in $u$, and canceling some terms, we find
\beq 
I_0 = \frac{1}{\gamma}{\rm arctan}\left(\frac{2\delta}{\gamma}\right)
+ \frac{3}{8}\left[2\delta - \gamma\,{\rm arctan}\left(\frac{2\delta}{\gamma}\right)\right]
= \frac{1}{\gamma}{\rm arctan}\left(\frac{2\delta}{\gamma}\right)
\left[1 + {\cal O}(\gamma^2)\right]~,
\eeq
\bea\label{I1}
I_1 & = & \frac{1}{2}\int_{-2\delta}^{2\delta} \frac{(1 - u^2/8)du}{u^2+\gamma^2}
\int_{-2\delta}^{2\delta} \frac{(1 + 3u^2/8)du}{u^2+\gamma^2}
- \frac{1}{2}\left(\int_{-2\delta}^{2\delta}\frac{du}{u^2+\gamma^2} \right)^2
\nonumber\\
& = & \frac{1}{8}\int_{-2\delta}^{2\delta} \frac{u^2 du}{u^2+\gamma^2}
\int_{-2\delta}^{2\delta} \frac{du}{u^2+\gamma^2}
= \frac{1}{2}\arctan \left(\frac{2\delta}{\gamma}\right)
\left[\left(\frac{2 \delta}{\gamma}\right) - 
\arctan \left(\frac{2\delta}{\gamma}\right)\right]
\left[1 + {\cal O}(\gamma^2)\right]~.
\nonumber\\
\eea

The values of $I_0$ for $\delta = (2\gamma, \gamma, \gamma/2)$ are respectively
$1.3258/\gamma, 1.1071/\gamma$, $\pi/(4\gamma)$, while those of $I_1$ 
are 1.7727, 0.4943 ,0.08427.
Thus we find 
\beq\label{I1/I0^2}
\frac{I_1}{(I_0)^2} =\left\{ \begin{array}{c c}  1.008\gamma^2 = 0.063(\Delta/m_\rho)^2 & 
~~~\delta = 2\gamma~~{\rm or}~~\Delta = 4\Gamma_\rho~, \cr
0.403\gamma^2 = 0.101(\Delta/m_\rho)^2& ~~~\delta = \gamma~~~~{\rm or}
~~\Delta = 2\Gamma_\rho~, \cr
0.137\gamma^2 = 0.137(\Delta/m_\rho)^2 & ~~\delta = \frac{1}{2}\gamma~~
{\rm or}~~\Delta = \Gamma_\rho~. \end{array}\right.
\eeq
\medskip
In the limit $\delta \ll \gamma$ 
(i.e., $\Delta \ll 2\Gamma_\rho$)
one may calculate $I_1/(I_0)^2$ using a Taylor expansion 
$\arctan(\delta/\gamma)= (\delta/\gamma) - (\delta/\gamma)^3/3 +
(\delta/\gamma)^5/5 - \ldots$. The dominant terms in $I_0$ and $I_1$ are 
\beq
I_0 = \frac{2\delta}{\gamma^2}~,~~~I_1 = \frac{\gamma}{6}\left(\frac{2\delta}{\gamma}\right)^3
\left(\frac{2\delta}{\gamma^2}\right) = \frac{8}{3}\frac{\delta^4}{\gamma^4}~,
\eeq
implying
\beq
\frac{I_1}{(I_0)^2} = \frac{2}{3}\delta^2
= \frac{1}{6}(\Delta/m_\rho)^2~.
\eeq
\medskip

\section{Two adjacent $\rho$ mass bands above and below $m_\rho$}

The longitudinal  decay rate $\Gamma_L(B^0 \to \rho^+\rho^-)$ is obtained by
integrating the amplitude squared over ranges $M_1$ for $m_{12}$ and $M_2$
for $m_{34}$ and {\em vice versa}, for two adjacent ranges $M_1$ and $M_2$
above and below $m_\rho$ situated  symmetrically with respect to $m_\rho$.
The interference term between $I=0,2$ and $I=1$ amplitudes is antisymmetric 
in $m_{12} \leftrightarrow m_{34}$ and vanishes when integrating these two
variables  over the two ranges $M_1$ and $M_2$ symmetrically,
\beq
\hskip-1mm
\left(\int_{M_1}\hskip-4mm dm_{12}\hskip-2mm \int_{M_2}\hskip-4mm dm_{34} 
\hskip-1mm + \hskip-2mm\int_{M_2}
\hskip-4mm dm_{12}\hskip-2mm\int_{M_1}\hskip-4mm dm_{34}\hskip-2mm\right)
\hskip-1mm\frac{{\rm Re}[f_{I=0,2}(m_{12}, m_{34})f_{I=1}^*(m_{12}, m_{34})]}
{ |m^2_{12} - m^2_\rho + im_\rho\Gamma_\rho|^2
|m^2_{34} - m^2_\rho + im_\rho\Gamma_\rho|^2} = 0~.
\eeq
Using $f_{I=0,2} = a, f_{I=1} = c(m_{12} - m_{34})/m_\rho$, the ratio of their contributions to the 
integrated decay rate is
$$
{\cal R}_L \hskip-1mm = \hskip-1mm\frac{c^2}{a^2}\hskip-1mm\int_{M_1}\hskip-1mm\int_{M_2}
\hskip-1mm\frac{[(m_{12} - m_{34})/m_\rho]^2dm_{12}dm_{34}}
{[(m^2_{12} - m^2_\rho)^2 + (m_\rho\Gamma_\rho)^2] [m_{12} \to m_{34}]}
\large/\hskip-2mm\int_{M_1}\hskip -1mm \frac{dm_{12}}
 {(m_{12}^2 -  m^2_\rho)^2 + (m_\rho\Gamma_\rho)^2}\times\hskip-1mm\small{\left[\hskip-2mm
 \begin{array}{c}  M_1 \to M_2 \cr
m_{12} \to m_{34} \end{array}
\hskip-2mm \right]}
$$
\beq
\equiv \frac{c^2}{a^2}\frac{{\cal I}_1}{{\cal I}_{01}{\cal I}_{02}}~.
~~~~~~~~~~~~~~~~~~~~~~~~~~~~~~~~~~~~~~~~~~~~~~~~~~~~~~~~~~~~~~~~~~~~~~
~~~~~~~~~~~~~~~~~~~~
\eeq
The double integral in the numerator denoted by ${\cal I}_1$ involves variables $m_{12}$ 
and $m_{34}$ which are larger and smaller than $m_\rho$, respectively.  
The two single-variable integrals in the denominator corresponding to the ranges $M_1$ and 
$M_2$ are denoted by ${\cal I}_{01}$ and ${\cal I}_{02}$, respectively. 

We now have
\beq
{\cal I}_{01} \equiv \int_1^{1+\delta}\frac{dx}{(x^2 - 1)^2 + \gamma^2}~,~~~~
{\cal I}_{02} \equiv  \int_{1 - \delta}^1\frac{dx}{(x^2 - 1)^2 + \gamma^2}~. 
\eeq
Substituting $u = x^2 -1, du = 2xdx = 2\sqrt{1+ u}\,dx$, expanding $(1 + u)^{-1/2} = 
1 - u/2 + \ldots$ and using the integrals (\ref{integrals}) and
\beq
\int^u\frac{udu}{u^2 + \gamma^2} = \half\ln(u^2 + \gamma^2)~,
\eeq
we obtain
\bea
{\cal I}_{01} & = & \frac{1}{2}\int_0^{2\delta}\frac{(1 + u)^{-1/2}du}
{u^2 + \gamma^2}
= \half\int_0^{2\delta}\frac{du}{u^2 + \gamma^2} - \frac{1}{4}\int_0^{2\delta}
\frac{udu}{u^2 + \gamma^2} 
\nonumber\\
& = & \frac{1}{2\gamma}\arctan\left(\frac{2\delta}{\gamma}\right)
- \frac{1}{8}\ln\left(\frac{4\delta^2 + \gamma^2}{\gamma^2}\right)~.
\eea
Similarly
\beq
{\cal I}_{02} = \frac{1}{2}\int_{-2\delta}^0\frac{(1 + u)^{-1/2}du}
{u^2 + \gamma^2} = \frac{1}{2\gamma}\arctan\left(\frac{2\delta}{\gamma}\right)
+ \frac{1}{8}\ln\left(\frac{4\delta^2 + \gamma^2}{\gamma^2}\right)~.
\eeq
That is, the leading terms in ${\cal I}_{01}$ and ${\cal I}_{02}$ behaving
like $1/\gamma$ are equal to each other. The subleading term occurring with
opposite signs affects the product ${\cal I}_{01}{\cal I}_{02}$ merely by its
square. We will neglect this correction of order $\gamma^2$ as we have done in
Sec.\,2:
\beq
{\cal I}_{01}{\cal I}_{02} = \left[\frac{1}{2\gamma}
\arctan\left(\frac{2\delta}{\gamma}\right)\right]^2[1 + {\cal O}(\gamma^2)]~.
\eeq

We now calculate ${\cal I}_1$ using binomial expansions as in Sec. 2: 
\bea\label{calI1}
{\cal I}_1 &\equiv & \int_1^{1+ \delta}\int_{1-\delta}^1\frac{(x_1-x_2)^2dx_1dx_2}
{[(x_1^2 - 1)^2 + \gamma^2][(x_2^2 - 1)^2 + \gamma^2]} 
 = \int_1^{1 + \delta}\frac{x_1^2dx_1}{(x_1^2 - 1)^2 + \gamma^2}
\int_{1- \delta}^1\frac{dx_2}{(x_2^2 - 1)^2 + \gamma^2}
\nonumber\\
& + &  \int_1^{1 + \delta}\frac{dx_1}{(x_1^2 - 1)^2 + \gamma^2}
\int_{1- \delta}^1\frac{x_2^2dx_2}{(x_2^2 - 1)^2 + \gamma^2}
-  2\int_1^{1 + \delta}\frac{x_1dx_1}{(x_1^2 - 1)^2 + \gamma^2}
\int_{1- \delta}^1\frac{x_2dx_2}{(x_2^2 - 1)^2 + \gamma^2}
\nonumber\\
& \simeq &  \hskip-1mm\frac{1}{8}\hskip-1mm\int_0^{2\delta}\hskip-1mm
\frac{du}{u^2+\gamma^2}
\hskip-2mm\int_0^{2\delta}\hskip-1mm\frac{u^2du}{u^2+\gamma^2}
+ \hskip-2mm\frac{1}{8}\left(\hskip-1mm\int_0^{2\delta}\hskip-1mm
\frac{udu}{u^2+\gamma^2}\right)^2
\nonumber\\
& = & \frac{1}{4}\hskip-1mm\left(\frac{1}{2}\arctan\left(\frac{2\delta}{\gamma}\right)
\left[\left(\frac{2\delta}{\gamma}\right) - \arctan\left(\frac{2\delta}{\gamma}\right)\right]
+ \frac{1}{8}\left[\ln\left(\frac{4\delta^2 + \gamma^2}{\gamma^2}\right)\right]^2\right)
\hskip-1mm\left[1 + {\cal O}(\gamma^2)\right].
\eea
Note that the first term in ${\cal I}_1$ equals $\frac{1}{4}I_1$ calculated in (\ref{I1}) in 
Sec.\,1. By itself it would have implied ${\cal I}_1/({\cal I}_{01}{\cal I}_{02}) = I_1/(I_0)^2$ 
because ${\cal I}_{01}{\cal I}_{02} = \frac{1}{4}I_0$.  
The additional $\ln$-squared term leads to a contribution with the same positive sign which 
is somewhat smaller than the first term for $\delta$ of order $\gamma$.  

The values of $\sqrt{{\cal I}_{01}{\cal I}_{02}}$ for $\delta = (2\gamma, \gamma, 
\gamma/2)$ are respectively
$0.6629/\gamma, 0.5536/\gamma$, $\pi/(8\gamma)$, while those of ${\cal I}_1$ 
are $0.6940, 0.2045 , 0.03608$.
Thus we find 
\beq
\frac{{\cal I}_1}{{\cal I}_{01}{\cal I}_{02}} =\left\{ \begin{array}{c c}  
1.579\gamma^2 & ~~~\delta = 2\gamma~, \cr
0.667\gamma^2 & ~~\delta = \gamma~, \cr
0.234\gamma^2 & ~~~\delta = \frac{1}{2}\gamma~. \end{array}\right.
\eeq
Comparing these results with (\ref{I1/I0^2}) and the line above we conclude
that taking two adjacent $\rho$ bands instead of a single common band leads to
suppression by a factor two of the dominant $I=0,2$ contribution and to
moderate enhancement of $57 - 71\%$ in the relative $I=1$ contribution.  

\section{Two separated $\rho$ meson bands each of width $\Gamma_\rho$}

In order to increase considerably the $I= 1$ contribution to the decay rate
relative to the $I=0, 2$ contribution we choose $M_1$ and $M_2$ to be two
$\rho$ bands each of width $\Gamma_\rho$, separated from each other by mass
ranges of width $\Gamma_\rho$ or $2\Gamma_\rho$.  These two cases correspond
to the following ranges in $x$:
\beq\label{ax}
{\rm (a)}~~1 - \frac{3}{2}\gamma \le x \le 1 - \frac{1}{2}\gamma,~~~~1 + \frac{1}{2}\gamma \le x 
\le 1 + \frac{3}{2}\gamma~,
\eeq
\beq\label{bx}
{\rm (b)}~~1 - 2\gamma \le x \le 1 - \gamma,~~~~~~~~~1 + \gamma \le x \le 1 + 2\gamma~.
\eeq
We study these two cases separately.

(a) Using notation and calculations as in Sec. 2 we then obtain
\beq
{\cal I}_{01} \equiv \int_{1+\gamma/2}^{1+3\gamma/2}\frac{dx}{(x^2 - 1)^2
 + \gamma^2}
= \frac{1}{2}\int_\gamma^{3\gamma}\frac{(1 + u)^{-1/2}du}{u^2 + \gamma^2}
=  \frac{1}{2\gamma}\left[\arctan(3) - \arctan(1)\right] - \frac{1}{8}\ln(5)~,
\eeq
\bea
{\cal I}_1 & \simeq &  \frac{1}{8}\int_{\gamma}^{3\gamma}\frac{du}{u^2+\gamma^2}
\int_{\gamma}^{3\gamma}\frac{u^2du}{u^2+\gamma^2}
+ \frac{1}{8}\left(\int_{\gamma}^{3\gamma}\frac{udu}{u^2+\gamma^2}\right)^2
\nonumber\\
& = & \frac{1}{8}\left[\arctan(3) - \arctan(1)\right]
\left[2 - \arctan(3) + \arctan(1)\right] + \frac{1}{32}\left[\ln(5)\right]^2\,,
\eea
implying
\beq\label{a} 
\sqrt{{\cal I}_{01}{\cal I}_{02}} = \frac{0.2318}{\gamma}~,~~~~{\cal I}_1 =
0.1700~,~~~ \frac{{\cal I}_1}{{\cal I}_{01}{\cal I}_{02}} = 3.16\gamma^2~.
\eeq

(b) For this range we calculate
\beq
{\cal I}_{01} \equiv \int_{1+\gamma}^{1+2\gamma}\frac{dx}{(x^2 - 1)^2 +
 \gamma^2}
= \frac{1}{2}\int_{2\gamma}^{4\gamma}\frac{(1 + u)^{-1/2}du}{u^2 + \gamma^2}
=\frac{1}{2\gamma}\left[\arctan(4) - \arctan(2)\right] - \frac{1}{8}\ln(17/5)~,
\eeq
\bea
{\cal I}_1 & \simeq &  \frac{1}{8}\int_{2\gamma}^{4\gamma}\frac{du}
{u^2+\gamma^2} \int_{2\gamma}^{4\gamma}\frac{u^2du}{u^2+\gamma^2}
+ \frac{1}{8}\left(\int_{2\gamma}^{4\gamma}\frac{udu}{u^2+\gamma^2}\right)^2
\nonumber\\
& = & \frac{1}{8}\left[\arctan(4) - \arctan(2)\right]
\left[2-\arctan(4)+\arctan(2)\right] + \frac{1}{32}\left[\ln(17/5)\right]^2\,,
\eea
implying
\beq\label{b}
\sqrt{{\cal I}_{01}{\cal I}_{02}} = \frac{0.1093}{\gamma}~,~~~~{\cal I}_1 =
0.0955~,~~~ \frac{{\cal I}_1}{{\cal I}_{01}{\cal I}_{02}} = 7.99\gamma^2~.
\eeq
The values of ${\cal I}_1/{\cal I}_{01}{\cal I}_{02}$ in (\ref{a}) and
(\ref{b}) should be compared with the much smaller value, $I_1/(I_0)^2 =
0.137\gamma^2$, obtained for this ratio for a common central $\rho$ mass band
of width $\Gamma_\rho$. The separation of the two $\rho$ bands by gaps
$\Gamma_\rho$ and $2\Gamma_\rho$ enhances the relative $I=1$ contribution by
factors of 23 and 58, respectively. These large enhancements are partially due
to a suppression of the $I=0,2$ contribution by factors of 11.5 and 52,
respectively.  

\section{Summary and conclusions} 

Let us compare overall decay rates and relative $I=1$ contributions to decay
rates for $\rho$ meson bands of decreasing width, considering first common
$\rho$ bands and then two separated bands for the two pairs of pions. We will
refer specifically to relevant measurements by the Babar and Belle
Collaborations, using for the $\rho $ mass and width the values\,\cite{PDG}
$m_\rho = 775$ MeV, $\Gamma_\rho = 148.5$ MeV implying $\gamma^2 = 0.0367$.

The Babar \cite{Aubert:2007nua} and Belle \cite{Vanhoefer:2015ijw}
collaborations studied longitudinally polarized $B^0 \to \rho^+\rho^-$ for the
two pion pairs forming a common $\rho$ band roughly of width $4\Gamma_\rho$,  
\beq 
m_\rho - 2\Gamma_\rho = 478~{\rm MeV}~\le m(\pi\pi) \le 1072~{\rm MeV} = 
m_\rho + 2\Gamma_\rho~.
\eeq
BaBar used a similar $\rho$ band for studying $B^+ \to \rho^+\rho^0$\,
\cite{Aubert:2009it}.  The averaged relative errors in the two measured decay
rates are around $7-8\%$\,\cite{Gronau:2016idx}.  The $I=1$ contribution to the
decay rate is characterized for this band by a quantity which 
is about half this error, 
\beq
\left[\frac{I_1}{(I_0)^2}\right]_{4\Gamma_\rho} = (1.008)(0.0367) = 0.037~.
\eeq

The Belle collaboration has measured $B^+\to \rho^+\rho^0$\,\cite{Zhang:2003up}
using a narrower band approximately of width $2\Gamma_\rho$, 
\beq 
m_\rho - \Gamma_\rho = 626~{\rm MeV}~\le m(\pi\pi) \le 924~{\rm MeV} =
m_\rho + \Gamma_\rho~.
\eeq
The measured decay rate involved a rather larger error (around $25\%$) because
the Belle analysis was based on only about ten percent of the final Belle
$\Upsilon(4S)$ sample.  Using our result
\beq
\frac{\left[(I_0)^2\right]_{2\Gamma_\rho}}{\left[(I_0)^2\right]_{4\Gamma_\rho}} = 0.70~,
\eeq
one expects with the complete Belle data sample an error in the decay rate
around $10\%$, somewhat larger than measured for a band of width
$4\Gamma_\rho$. The relative $I=1$ contribution to the decay rate expected for
a band of width $2\Gamma_\rho$, characterized by  
\beq
\left[\frac{I_1}{(I_0)^2}\right]_{2\Gamma_\rho} = (0.403)(0.0367) = 0.015~,
\eeq
is about $40\% \simeq 0.015/0.037$ of the one for a band of width
$4\Gamma_\rho$.  In order to reach this sensitivity in measurements of $B \to
\rho\rho$ decay rates one needs about $6 \simeq (0.037/0.015)^2$ times more
data than accumulated so far.

For an even narrower $\rho$ band of width $\Gamma_\rho$, as studied briefly 
by Belle~\cite{Vanhoefer:2015ijw}, 
\beq
m_\rho - \half\Gamma_\rho = 701~{\rm MeV}~\le m(\pi\pi) \le 849~{\rm MeV} =
m_\rho + \half\Gamma_\rho~,
\eeq
we calculate
\beq \label{eqn:gam4gam}
\frac{\left[(I_0)^2\right]_{\Gamma_\rho}}{\left[(I_0)^2\right]_{4\Gamma_\rho}} = 0.35~.
\eeq
Thus for this range one requires about three times more data than used by Babar
and Belle for measuring an error of $8\%$ on $B\to \rho\rho$ decay rates, and
about fifty times more data for reaching an accuracy of two percent in these
rates (or one percent in corresponding amplitudes).
[That is 1/4 of the present 8\%, requiring 16 times more data, and
multiplying 16 by the factor of three mentioned just below Eq.\
(\ref{eqn:gam4gam})].
The $I=1$ contribution for such a narrow band is characterized by an even
smaller number:
\beq\label{Gamma}
\left[\frac{I_1}{(I_0)^2}\right]_{\Gamma_\rho} = 0.005~.
\eeq
Therefore, unless $c^2/a^2$ is considerably larger than one, measuring $B\to
\rho\rho$ decay rates for a $\rho$ meson band of width $\Gamma_\rho$ with
fifty times more data than used by Babar and Belle is expected to yield
values for $B\to \rho\rho$ amplitudes which are insensitive to the $\rho$
width. Such a data sample is expected at the SuperKEKB Belle II experiment
\cite{Abe:2010gxa,Aushev:2010bq}.

If $c^2/a^2 \gg 1$ the $I=1$ contribution to the decay rate for a $\rho$ band
of width $\Gamma_\rho$ is considerably larger than one percent. In this case
one would hope to be able to subtract this contribution from the measured decay
rate in order to obtain the pure $I=0, 2$ contribution. This would require
a higher sensitivity to the $I=1$ contribution.  For this purpose we have
studied two pairs of pions for two separated ranges of dipion masses. 

For two $\rho$ bands each of width $\Gamma_\rho$ separated by a range of width
$\Gamma_\rho$, described by Eqs.\ (\ref{ax}) and (\ref{a}),
\beq 
m_\rho - \frac{3}{2}\Gamma_\rho = 552~{\rm MeV}~\le m(\pi\pi) \le 701~{\rm MeV}
= m_\rho - \frac{1}{2}\Gamma_\rho~,
\eeq
\beq 
m_\rho + \frac{1}{2}\Gamma_\rho = 849~{\rm MeV}~\le m(\pi\pi) \le 998~{\rm MeV}
= m_\rho + \frac{3}{2}\Gamma_\rho~,
\eeq
we calculated using the above value of $\gamma^2$
\beq \label{eqn:gamratio}
\frac{{\cal I}_1}{{\cal I}_{01}{\cal I}_{02}} = (3.16)(0.0367) = 0.116~,
\eeq
while for two bands of width $\Gamma_\rho$ separated by a mass range
$2\Gamma_\rho$, described by Eqs.\ (\ref{bx}) and (\ref{b}),
\beq
m_\rho - 2\Gamma_\rho = 478~{\rm MeV}~\le m(\pi\pi) \le 626~{\rm MeV} = 
m_\rho - \Gamma_\rho~,
\eeq
\beq 
m_\rho + \Gamma_\rho = 924~{\rm MeV}~\le m(\pi\pi) \le 1072~{\rm MeV} = 
m_\rho + 2\Gamma_\rho~,
\eeq
one finds
\beq
\frac{{\cal I}_1}{{\cal I}_{01}{\cal I}_{02}} = (7.99)(0.0367) = 0.293~.
\eeq
These values are much larger than (\ref{Gamma}) calculated for a single $\rho$
band of width $\Gamma_\rho$, which indicates a higher sensitivity to the
$I=1$ amplitude.  
The effect of such an enhanced $I=1$ amplitude on the $B \to \rho \rho$ rate
would be apparent when comparing the branching fraction obtained from a single
band of width $\Gamma_\rho$ with that obtained from two bands of width
$\Gamma_\rho$ separated by a gap of $\Gamma_\rho$.
We have also shown large suppressions of the $I=0,
2$ contributions for two bands separated by $\Gamma_\rho$ and $2\Gamma_\rho$,
\beq
\frac{\left[{\cal I}_{01}{\cal I}_{02}\right]_{{\rm separated~by}~(\Gamma_\rho,
2\Gamma_\rho)}}
{\left[(I_0)^2\right]_{\Gamma_\rho}} = (\frac{1}{11.5}~,~~\frac{1}{52})~.
\eeq

Consider the first case (a) ``Separated by $\Gamma_\rho$" in the above
equation.  ${\cal I}_{01}{\cal I}_{02}/(I_0)^2 = 1/11.5$ implies that this
range requires about 33 = 11.5/0.35 times more data than used by Babar and 
Belle (using a range $4\Gamma_\rho$) for measuring an error of 8\% in the 
$B \to \rho \rho$ decay rate. Using 50 times more data, as expected at SuperKEKB, 
would reduce this error in decay rate to 
$8\%\sqrt{33/50} = 6.5\%$.

Now assume, for instance, $c^2/a^2 = 10$ in which case (\ref{eqn:gamratio})
implies $(c^2/a^2) {\cal I}_1/({\cal I}_{01}{\cal I}_{02}) = 1.16 = 116\%$
which is 18 times the above error of 6.5\%. Actually, in this case the decay
rate has two comparable contributions from $a^2{\cal I}_{01}{\cal I}_{02}$ and
$c^2{\cal I}_1$ and the error is smaller. 
Specifically, with $c^2/a^2 = 10$ the decay rate for case (a) is
$a^2[{\cal I}_{01}{\cal I}_{02} + 10 {\cal I}_1] = 2.16a^2 {\cal I}_{10}{\cal
I}_{20}$ which is $(2.16/11.5) [a^2(I_0)^2]_{\Gamma_\rho} =
[a^2(I_0)^2]_{\Gamma_\rho}/5.3$.  This then implies that this separated range
requires 15 = 5.3/0.35 times more data than used by Babar and Belle for an 8\%
error in decay rate.  Using 50 times more data than used by Babar and Belle
would reduce this error in the measured decay rate to $8\%/\sqrt{50/15}=8\%/1.8
= 4.3\%$, and a corresponding error of this order in the relatively large $I=1$
contribution.  Subtraction of the thus-determined $I=1$ contribution, 
$(c^2/a^2)[I_1/(I_0)^2]_{\Gamma_\rho}=5\%$ [cf.\ Eq.\ (\ref{Gamma})], from the
decay rate measured for a single band of width $\Gamma_\rho$ would then yield a
very high-precision value for the pure $I=0,2$ contribution in this decay, 
involving an error around $4.3\%\times 5\% = 0.2\%$ from uncertainty in the
$I=1$ amplitude.

To conclude, unless $c^2/a^2 \gg 1$, the best way to place limits on the $I=1$
amplitude seems to be the method of Sec.\ 2, i.e., by considering the two
$\rho$ mesons in a common band of width 
$\Delta \equiv 2m_\rho\delta$ centered at $m_\rho$, decreasing $\delta$ until limited by statistics.  
For the expected ratio $c^2/a^2 = {\cal O}(1)$, the extracted $B \to \rho \rho$ branching 
ratios should approach a constant value, with negligible $I=1$ contribution, as $\delta$
is decreased sufficiently down to $\Gamma_\rho/2m_\rho$ and below, 
namely when $\Delta \le \Gamma_\rho$.  
In contrast, if for some unknown reason $c^2/a^2 \gg 1$ then the branching ratio 
would decrease with decreasing values of $\delta$, as a nonnegligible positive $I=1$
contribution to the decay rate became smaller.
In this case the effect of the $I=1$ amplitude would show up as an artificial
enhancement of the $B \to \rho \rho$ decay rate 
measured for two $\rho$ bands of width $\Gamma_\rho$ separated by a gap
$\Gamma_\rho$ when compared with the decay rate obtained from a single band
of width $\Gamma_\rho$.  Translating the enhancement in the former decay to
the $I=1$ contribution in the latter would yield a very precise value for the
$I=0,2$ amplitude used in the isospin analysis.

\end{document}